\documentstyle[prl,aps,preprint]{revtex}

\begin{document}
\def\lsim{\buildrel <\over\sim }
 
\title{Coherent quasiparticle weight
and its connection to high-$T_c$ 
superconductivity from angle-resolved photoemission}
\author{
        H. Ding,$^1$
        J.R. Engelbrecht,$^1$ Z. Wang, $^1$
        J. C. Campuzano,$^{2,3}$
        S.-C. Wang, $^1$ H.-B. Yang, $^1$ R. Rogan,$^1$  
        T. Takahashi,$^4$
        K. Kadowaki,$^5$
        and D. G. Hinks $^3$
       }
\address{
         (1) Department of Physics, Boston College, Chestnut Hill, MA 
             02467 \\
         (2) Department of Physics, University of Illinois at Chicago,
             Chicago, IL 60607\\
         (3) Materials Sciences Division, Argonne National Laboratory,
             Argonne, IL 60439 \\
         (4) Department of Physics, Tohoku University, 980 Sendai, Japan\\
         (5) Institute of Materials Science, University of Tsukuba, 
             Ibaraki 305, Japan\\
         }

\maketitle
\bigskip
\bigskip

{\bf

In conventional superconductors, the pairing energy gap ($\Delta$)
and superconducting phase coherence go hand-in-hand. As the 
temperature is lowered, both the energy gap and phase coherence appear 
at the transition temperature $T_c$.  
In contrast, in underdoped high-$T_c$ superconductors 
(HTSCs), a pseudogap appears at a  much higher temperature $T^*$, 
smoothly evolving into the superconducting gap at $T_c$ 
\cite{NATURE1,LOESER}. Phase coherence on the other hand is only
established at $T_c$, signaled by the appearance of a sharp 
quasiparticle (QP) peak \cite{qpnote}
in the excitation spectrum. 
Another important difference between the two types of 
superconductors is in the ratio of $2\Delta / T_c\equiv R$. 
In BCS theory, $R\sim3.5$, is constant.
In the HTSCs this ratio varies widely, continuing 
to increase in the underdoped region, where the gap increases while 
$T_c$ decreases. 
Here we report that in HTSCs it is the ratio 
$z_{\it A}\Delta_m/T_c$ which is approximately constant, where $\Delta_m$ 
is the maximum value of the $d$-wave gap, and $z_{\it A}$ is the weight 
of the coherent excitations in the spectral function. 
This is highly unusual, since in nearly all phase transitions,
$T_c$ is determined by an energy scale alone.
We further show that in the low-temperature limit, $z_{\it A}$ increases 
monotonically with increasing doping $x$. The growth is linear, 
{\it i.e.} $z_{\it A}(x)\propto x$, in the underdoped to optimally 
doped regimes, and slows down in overdoped samples. The 
reduction of $z_{\it A}$ with increasing temperature resembles that of 
the $c$-axis superfluid density.
}

This brings us to the important question of the meaning of $z_{\it A}$ and
its determination by angle-resolved photoemission spectroscopy (ARPES). We
have recently shown \cite{KAMINSKIQP} that, although the ARPES spectral
function is very broad in the normal state, indicating that there are no
quasiparticles, in the superconducting state it separates into coherent
and
incoherent components everywhere along the Fermi surface. We call the
coherent component the quasiparticle piece and
its spectral weight (normalized energy integral), $z_{\it A}$.
For a Fermi liquid, this is the quasiparticle residue $z$. 
The validity of a Fermi liquid picture in the
superconducting state of HTSCs has not been established,
but the restoration of electronic coherence below $T_c$ is demonstrated by 
a developing $z_{\it A}$ in ARPES spectra.
In the highly anisotropic HTSCs one would expect $z_{\it A}$
to dependent significantly on the in-plane momentum.
We focus our analysis on the coherent portion of the spectral weight
in the vicinity of the $(\pi,0)$ point of the Brillouin zone.  
The vicinity of $(\pi, 0)$ contributes most of the angle-integrated
spectral weight. This assertion derives from a comparison of the density
of
states as measured by scanning tunneling microscopy (STM)\cite{DEWILDE}
and
the ARPES spectral function at $(\pi, 0)$ on identical samples, shown in
the
inset of Fig.~1(b). Once the tunneling spectrum is modified by a Fermi
function and convoluted with the ARPES energy resolution, we find a
remarkable similarity between the two, including the details of the
lineshape, giving us a good reason to believe that the spectral function in
the vicinity of $(\pi,0)$ dominates the total density of states.
In addition, the $d$-wave gap, and therefore the pairing energy scale,
is maximized at $(\pi,0)$.

In Fig.~1a we show the ARPES spectra \cite{EXPERIMENT} 
at $(\pi,0)$ for an optimally 
doped Bi2212 sample as a function of temperature. The evolution from 
a broad, incoherent spectral function at high temperature to one that 
has a sharp peak followed by a broad incoherent part at low temperatures 
can clearly be seen \cite{RANDERIA}. Note that a spectral loss (dip) 
also develops on the high binding energy side of the QP 
in the SC state, as compared to the normal state spectrum \cite{DIP}. 
The dip separates the coherent QP from the incoherent part 
(hump). Most of the intensity associated with the
incoherent hump is believed to be an intrinsic part of the 
single-particle spectral function, based on the observations that it has 
the same photoemission matrix element as the coherent QP \cite{FSPRL} 
and its position scales with that of the QP \cite{HUMPPRL}. Note that 
although the QP peak appears at $T_{c}$ in optimally doped samples, 
it is not a direct consequence of the formation of a gap, but is instead 
related to the onset of phase coherence \cite{NATURE1}, in sharp contrast 
to a  BCS superconductor.

We analyze the ARPES data in Fig.~1, by fitting a sharp Gaussian function
to the coherent peak and a broad Lorentzian with an asymmetric cutoff 
to the hump. Both are multiplied by a Fermi function. Although one 
would expect the quasiparticle peak to have a Lorentzian 
lineshape, we find that a Gaussian best fits the actual 
lineshape at low temperature. 
This remains  the case even for our high-resolution data 
(with a resolution (FWHM) as high as $7$ meV, not shown here), 
suggesting that the QP peak is not resolution limited -- consistent 
with a previous report \cite{JOHNSON}. A possible origin of such a
lineshape
is an averaging over a random distribution of a large number of
sharper peaks arising from inhomogeneities observed in the 
tunneling data by STM on Bi2212 \cite{PAN} which shows a 
Gaussian-like gap distribution with a width of $\sim20$ meV. 
It is reassuring that at 
higher temperatures the fit is consistent with a Lorentzian. The fit is 
insensitive to the form of the broad function describing the hump. Further 
fits to data over a wide range of doping values will be reported 
elsewhere \cite{ROGAN}.  

In Fig.~2 we give an example of how our fits separate the sharp QP from 
the incoherent spectrum of a slightly overdoped Bi2212 sample. 
From the Gaussian fit, we obtain the QP weight $z_{\it A}$, the QP 
line-width $\Gamma$, and  the QP peak position which gives the maximum gap 
$\Delta_m$. 
We obtain $z_{\it A}$ from the 
ratio of the area under the fitted QP peak to the area of the total 
energy distribution curve (EDC) integrated over the range: 
[$E_{\rm min},+\infty$], where $E_{\rm min}$ is the minimum of the 
EDC (below the hump), in the vicinity of $-0.5$ eV \cite{halfz}. 
The choice of integrating range is based on the assumption that 
$E_{\rm min}$ is where the conduction Cu-O band separates 
from other bands \cite{FSPRL}. 
Note that the overall amplitude of $z_{\it A}$ may be underestimated 
by this method because some of the extrinsic background (although small) 
is included in the denominator. However, we do not expect it to 
affect the doping and temperature dependence of $z_{\it A}$.

We first present our results at a fixed low temperature ($14$K).
Fig.~3 shows $z_{\it A}$ and $\Delta_m$ {\it v.s.} the doping
concentration $x$. From Fig.~3(a) we see that the QP weight grows 
{\em linearly} with $x$ in underdoped and optimally doped samples, 
and tapers off on the overdoped side. Together with the observation 
that the area enclosed by the normal state Fermi surface (obtained 
in ARPES as the locus of gapless excitations) scales as $1-x$ 
\cite{HONG}, the finding of $z_{\it A}\propto x$ suggests that only $x$ 
number of coherent carriers are recovered in the SC state, 
consistent with the picture
of doping a Mott insulator with $x$ holes.
The maximum superconducting gap $\Delta_m$ at $(\pi,0)$ is plotted in 
Fig.~3(b) as the QP peak position \cite{DINGJPCS}. This plot shows a 
trend that $\Delta_m$ increases linearly 
with decreasing doping in contrast to the behavior of $T_c$.

We next look at the reduction of the QP coherence upon heating, shown in 
Fig.~4, where we plot $z_{\it A}(T)$, $\Delta_m(T)$ and $\Gamma(T)$ for
three 
typical samples in the underdoped, optimally doped, and overdoped regions.
At optimal doping (Fig.~4(a)), $z_{\it A}(T)$ is only weakly $T$-dependent
at 
low temperatures, but falls off dramatically as $T$ is increased towards 
$T_c$, which is consistent with the qualitative trend reported in previous 
ARPES studies \cite{RANDERIA,LOESER2,JOHNSON}. The overall temperature
dependence of $z_{\it A}(T)$ remarkably resembles that of the $c$-axis
superfluid
density measured by Josephson plasma resonance in Bi2212 
\cite{GAIFULLIN}, penetration depth in YBCO \cite{BONN}, and ac 
susceptibility in LSCO \cite{PANA1}. A possible explanation is that the 
interlayer tunneling matrix element is enhanced near the $(\pi,0)$ point 
\cite{interlayer} such that the low-temperature QP weight near $(\pi,0)$ 
contributes substantially to the $c$-axis superfluid density.
An interesting recent theoretical work \cite{CARLSON} within the stripe
picture predicts a direct relation between the weight of the QP peak 
and the $ab$-plane superfluid density. However, the connection we find 
here is between the QP weight at $(\pi,0)$ and the $c$-axis superfluid 
density which has a much weaker temperature
dependence than the $ab$-plane superfluid density at low temperature. 
In Fig.~4(b) we compare the $T$-dependence of $z_{\it A}(T)/z_{\it A}(0)$ 
between an underdoped and an overdoped sample. The coherent 
weight drops faster in the overdoped sample at low temperatures, 
once again reminiscent of the $c$-axis superfluid density in overdoped 
LSCO and the trend that the $c$-axis superfluid density is depleted 
faster upon heating as doping is increased \cite{PANA1}.

In all three cases, $z_{\it A}$ drops significantly as $T$ approaches
$T_c$.
However, the extracted value of $z_{\it A}$ is nonzero above, but close to
$T_c$. 
We caution that the error bars increase significantly above $T_c$ where
the overall contribution from the QP becomes very small and thus difficult
to 
separate from the incoherent part of the spectrum. Taking a nonzero
$z_{\it A}$ 
above $T_c$ at face value might suggest that the QP has already formed
above $T_c$ for all doping values. However, a closer look at the
temperature
dependence of the QP position and line-width plotted in Fig.~4(c)
shows that there are qualitative differences between the underdoped and 
overdoped regimes near $T_c$. In all cases the QP line-width saturates 
at low temperatures due to inhomogeneities described above. Thus the 
low-temperature line-width should not be regarded as the intrinsic QP 
scattering rate which might be much smaller. However, for the 
underdoped sample, the line-width increases quite rapidly
with increasing temperature while the QP position remains roughly
unchanged until the width crosses the position near $T_c$, and
the QP loses its identity. The opposite trend is found for
the overdoped sample. Here the width remains approximately independent
of $T$ across $T_c$ while the position decreases. Thus, it is the loss 
of coherence near $T_c$ that destroys the QP on the underdoped side 
\cite{NORMAN}, but the closing of the energy gap near $T_c$ that weakens 
the QP signature above $T_c$ on the overdoped side.

A natural conclusion is that superconducting order is established 
through an emerging QP coherence $z_{\it A}$ in the underdoped regime 
(where $\Delta_m\ne0$ above $T_c$), while it is controlled by the 
development of the superconducting gap $\Delta_m$ on the overdoped 
side. This, on the gross level, is consistent with the original 
resonating valence bond picture \cite{andersonscience} and its 
variants \cite{RVB2}. Motivated by our results, we conjecture that 
a new quantity $z_{\it A}(0)\Delta_m$, with the dimension of energy, 
possibly plays the role of the superconducting order parameter and  
determines $T_{c}$. In Fig.~5, we plot $z_{\it A}(0)\Delta_m$ and $T_c$ 
{\it vs} $x$, which reveals a striking proportionality between 
the two quantities \cite{zdeltanote}. We conclude that for Bi2212,
$$
R={z_{\it A}(0)\Delta_m\over k_BT_c}={\rm constant},
$$
as demonstrated in the inset of Fig.~5. This result differs from the 
BCS theory. It is known that the
effect of $z_{\it A}$ typically does not enter this formula in the 
Fermi liquid approach. The experimental findings reported here 
strongly suggest that, unlike in conventional superconductors, single 
particle coherence plays an important role in high-$T_c$
superconductivity. 
It is interesting to note that the relation $x\Delta_m/k_BT_c\approx 3J/t$ 
was derived recently in the {\it underdoped} regime in 
a gauge theory formulation of the $t-J$ model \cite{leewen}. This is 
consistent with our observations provided that 
$z_{\it A}\propto x$ holds in this theory.

\vfill\eject

We thank T. Yokoya, T. Takeuchi, A. Kaminski, T. Sato, H. Fretwell, J. 
Mesot, Y. DeWilde for their experimental help and M.R. Norman for
theoretical discussion. We also thank S.H. Pan for providing us 
unpublished STM results. HD is supported by the Sloan Research fellowship.
This work is supported by US NSF and DOE,the CREST of JST, 
and the Ministry of Education, Science and Culture of Japan.
The Synchrotron Radiation Center is supported by the NSF.

Correspondence and requests for materials should be addressed to H.~D.~ 
(e-mail: dingh@bc.~edu)

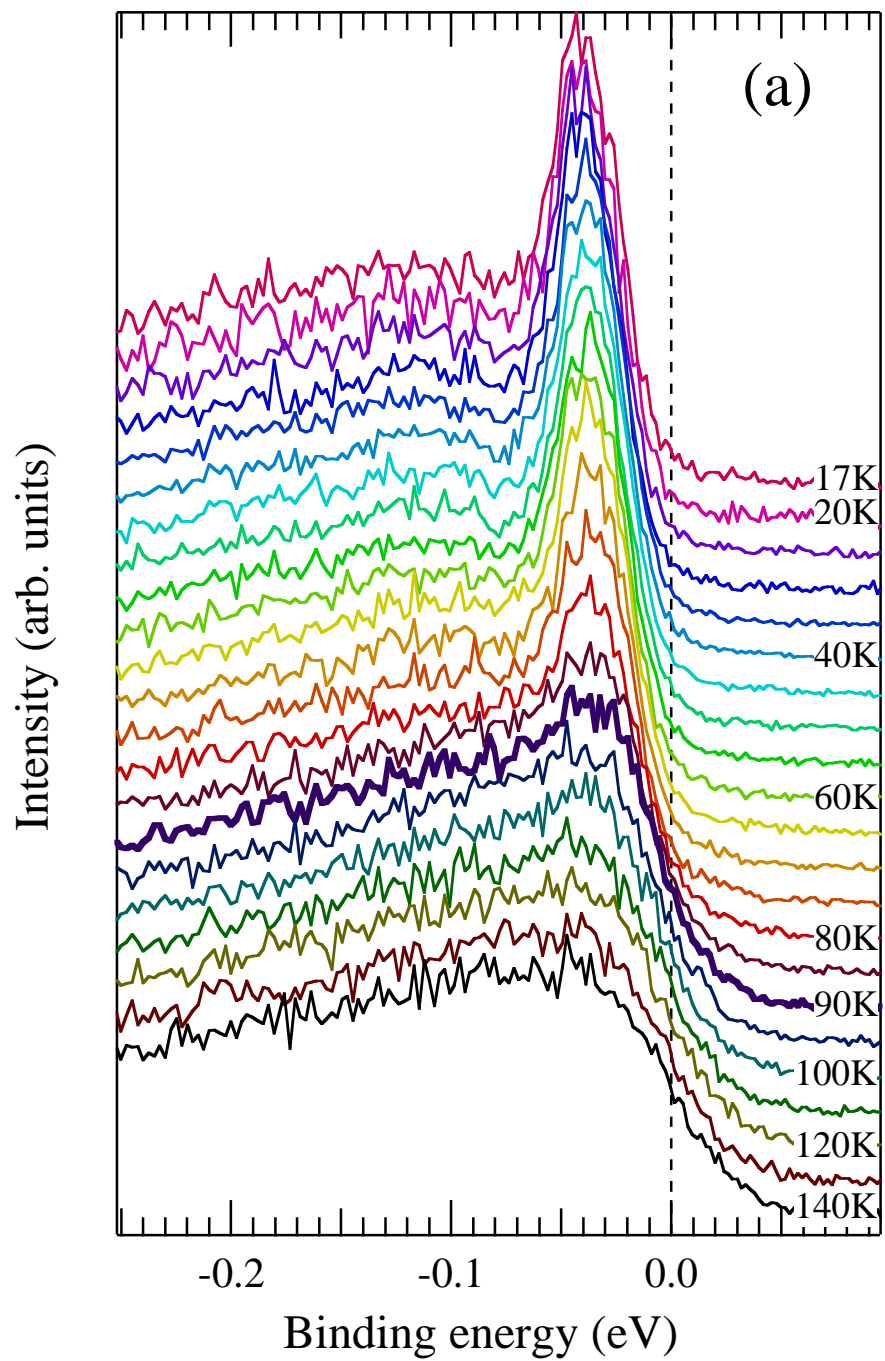
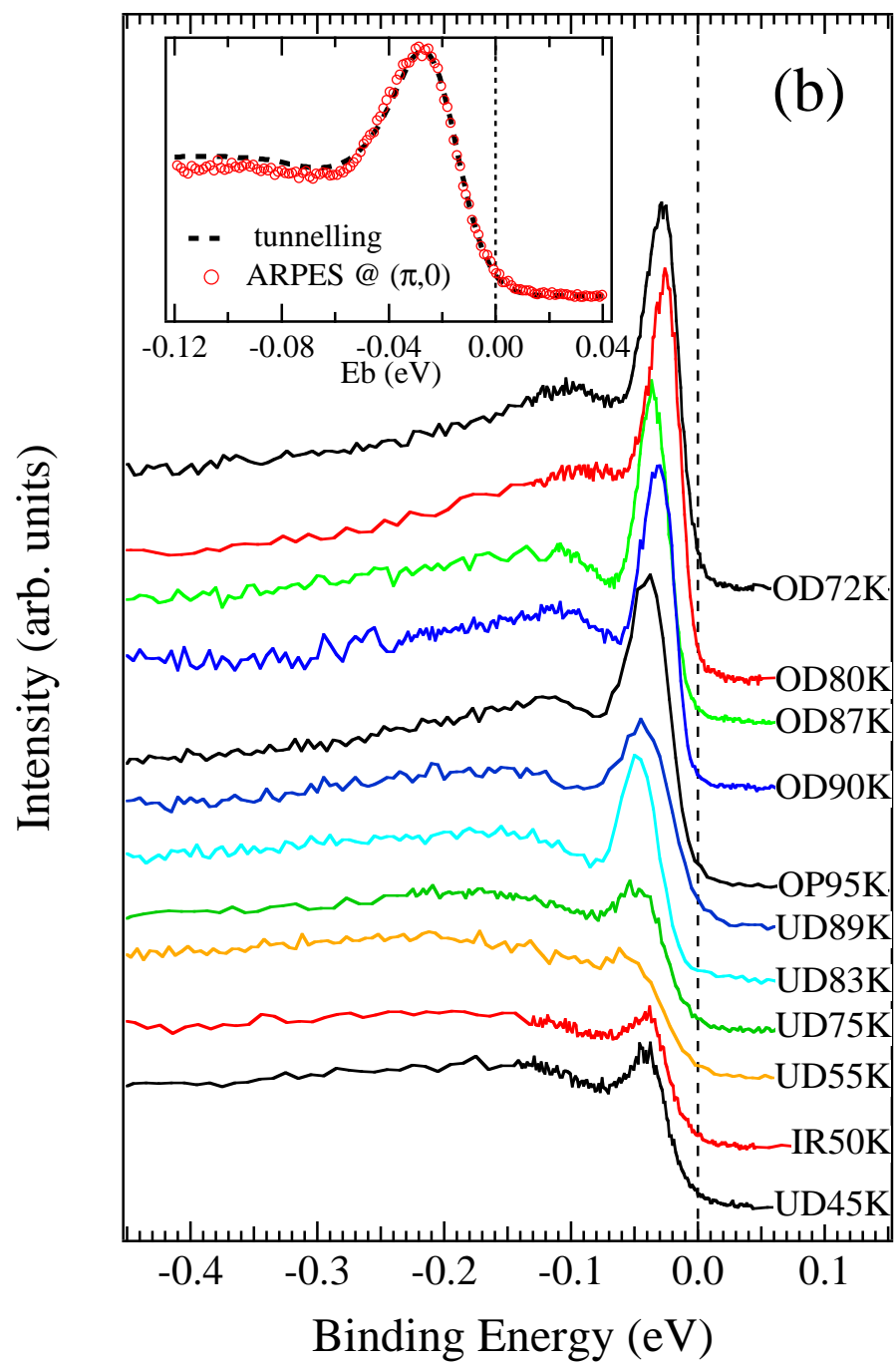
\begin{figure}
\caption{
(a) ARPES spectra at $(\pi,0)$ of slightly overdoped doped Bi2212
($T_c$=90K) for different temperatures ($T=$
17,20,25,30,35,40,45,50,55,60,65,70,75,80,85,90,95,100,110, 
120,130,140K 
from top to bottom).   (b) Spectra at $(\pi,0)$ at low 
temperature ($T=14$K) of differently doped Bi2212 samples (OD - 
overdoped, OP - optimally doped, UD - underdoped, IR - 300 MeV electron 
irradiated, followed by the value of $T_c$).  Spectra intensity are 
normalized at a high binding energy where the spectral intensity shows a
minimum (in the vicinity of -0.5 eV.). Inset: Comparison between 
low-temperature ARPES at $(\pi,0)$ and STM for the same OD72K sample. 
}
\label{fig1}
\end{figure}

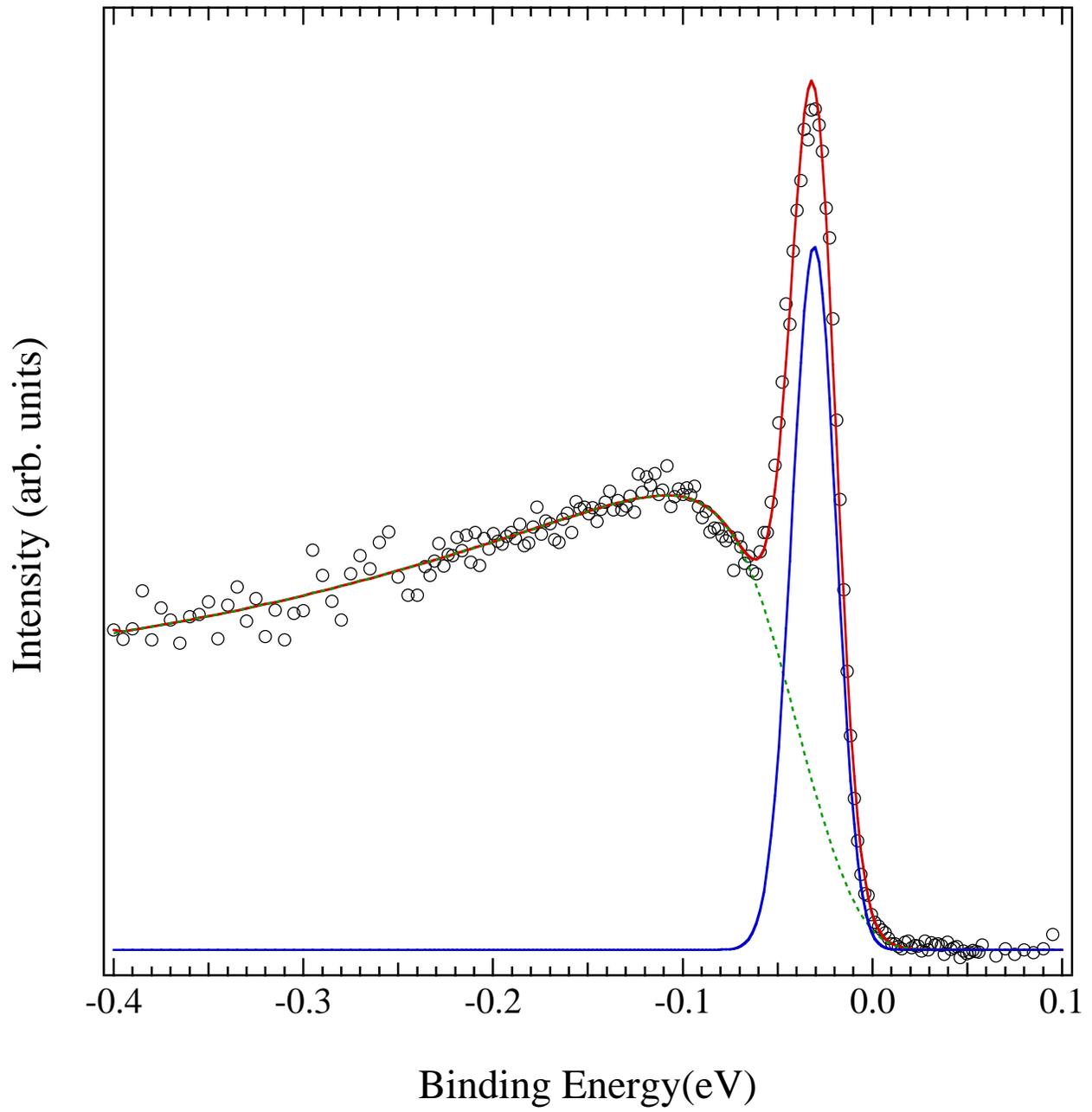
\begin{figure}
\caption{
A fitting example of a low-temperature (14K) spectrum (black open circles)
of 
slightly
overdoped Bi2212 ($T_c$=90K) at $(\pi,0)$. The blue solid line is a sharp 
Gaussian representing the coherent peak. The green dashed line is a broad 
Lorentzian cut by an asymmetric cutoff function for the 
incoherent part. The sum of the coherent and the incoherent part gives 
the fitting result (red solid line).
}
\label{fig2}
\end{figure}

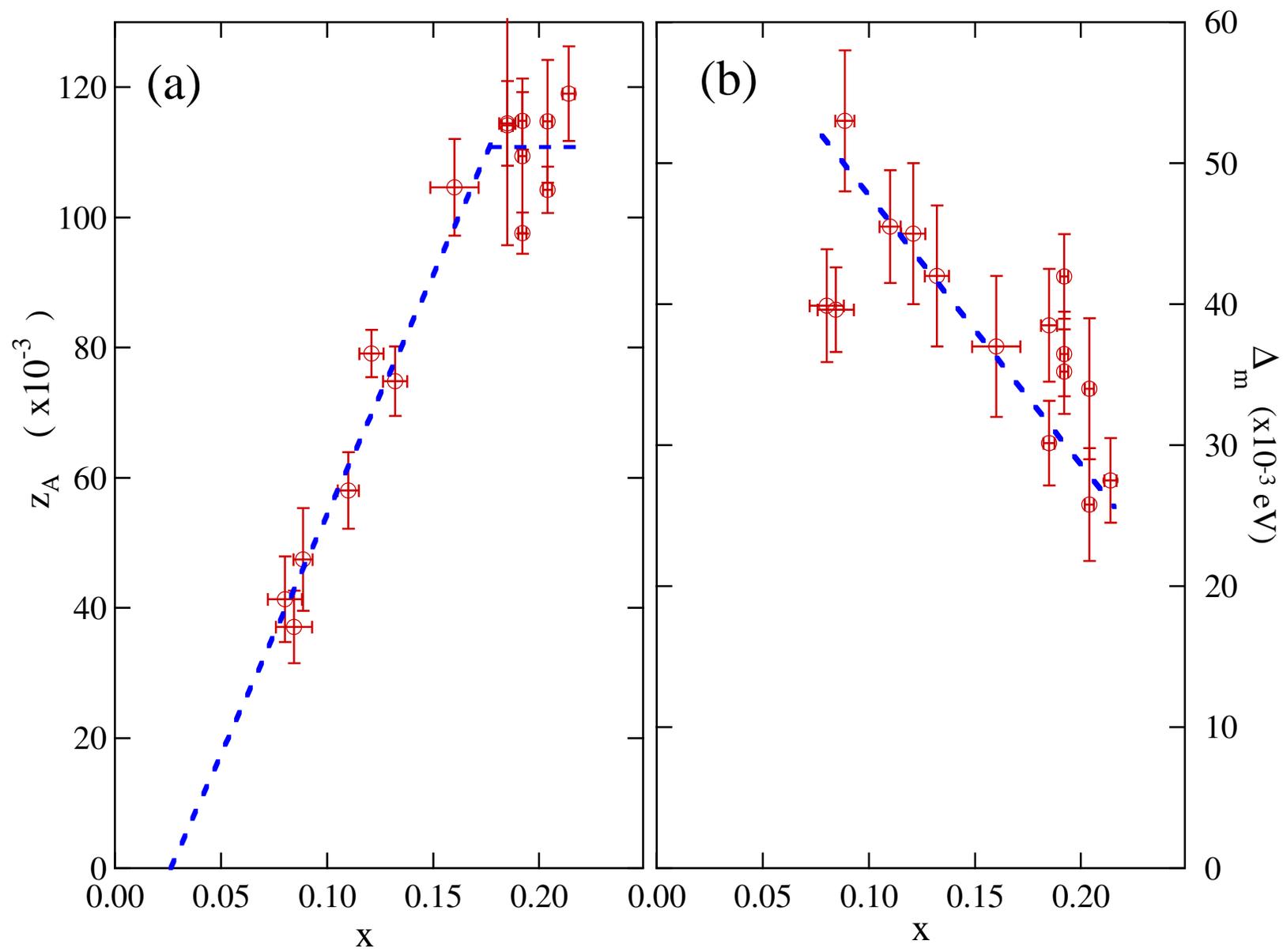
\begin{figure}
\caption{
(a) Doping dependence of the low-temperature (14K) 
coherent spectral weight 
$z_{\it A}$.  The dash line is a guide line showing that $z_{\it A}$
increases 
linearly on underdoped side, and tapers off on the overdoped side. 
(b) Doping dependence of the maximum gap size $\Delta_m$ at 14K
obtain from the position of the coherent QP peak from our fit 
procedure. Vertical error bars plotted in this and following figures 
are mostly from fitting uncertainty rather than from measurement. 
Notice that two heavily underdoped (UD45K and IR50K) has smaller gaps. 
This may be due to the effect of impurities as reflected in their 
broader transition width. 
}
\label{fig3}
\end{figure}

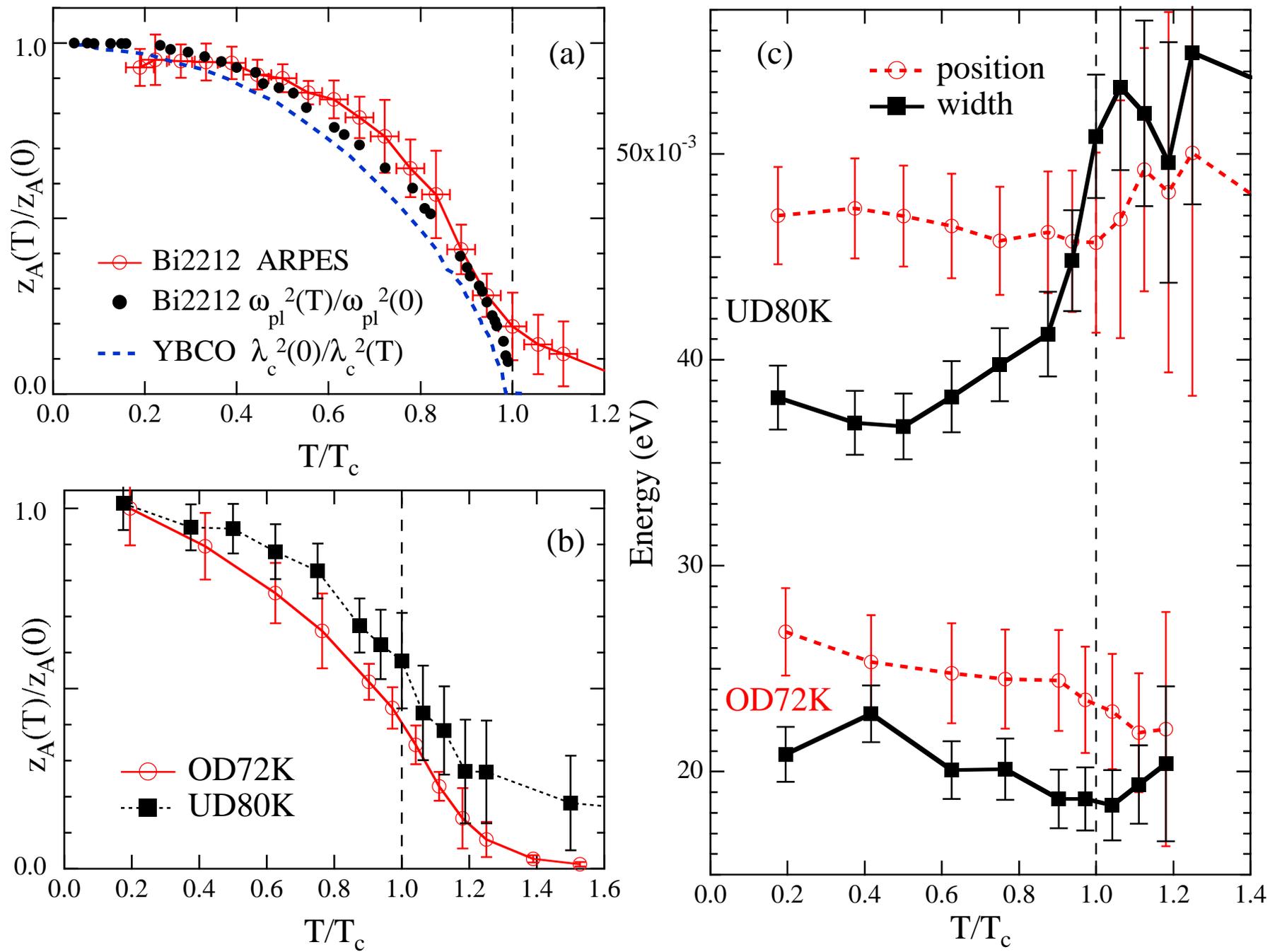
\begin{figure}
\caption{
Temperature-dependence of the extracted QP properties for three samples 
(OD72K, OD90K, and UD80K) near $(\pi,0)$.  
(a) normalized $z_{\it A}(T)/z_{\it A}(0)$ vs $T/T_c$ for OD90K Bi2212
compared with
normalized $c$-axis superfluid density obtained from 
Josephson plasma resonance \protect\cite{GAIFULLIN} 
of overdoped Bi2212 ($T_c=82$K) and microwave penetration 
depth measurements 
\protect\cite{BONN} of optimally doped YBCO ($T_c=93.5$K).
(b) Normalized QP weight, $z_{\it A}(T)/z_{\it A}(0)$ v.s. $T/T_c$,
comparing OD72K and UD80K samples.
(c) QP position (that defines $\Delta_m$) and QP width v.s. $T/T_c$,
again comparing OD72K and UD80K samples.  
The effect of the energy resolution ($\sim$15 meV) 
is removed from the line-width through the approximate relation 
$\Gamma=\sqrt{\Gamma_{\rm measured} ^2 - {\rm Resolution}^2}$.  
}
\label{fig4}
\end{figure}

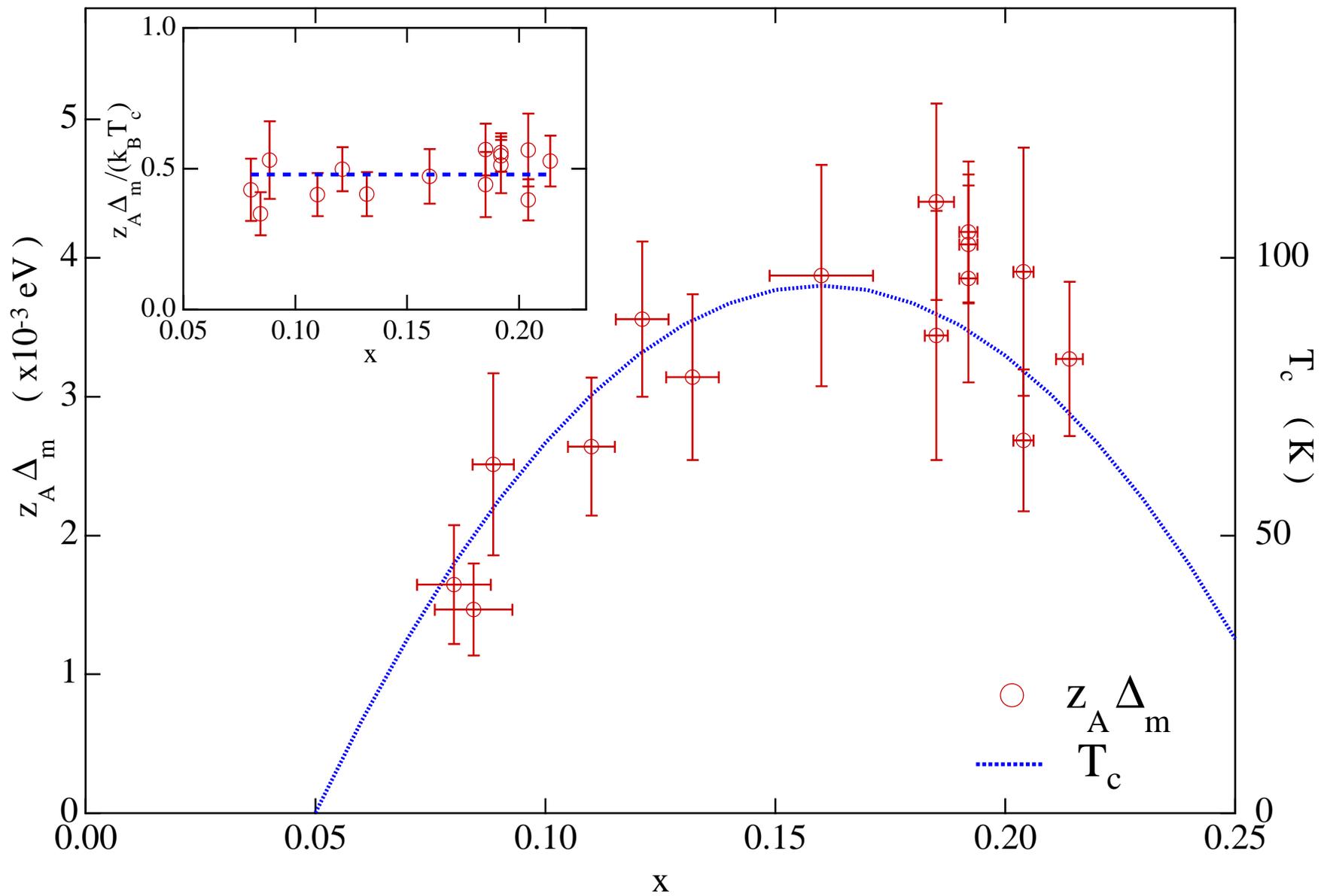
\begin{figure}
\caption{
Doping dependence of the value of $z_{\it A}\Delta_{m}$  (open circles) at 
$(\pi, 0)$ at low temperature $(T=14K)$.
The dash line is the empirical relation 
\protect\cite{tcvsx}
between $T_c$  and $x$ given by 
$T/T_c^{\rm max}=1-82.6(x-0.16)^2$ with 
$T_c^{max}$=95K. The inset shows that the ratio of $z_{\it A}\Delta_m$ and 
$k_BT_c$ is a constant over the doping range studied.}
\label{fig5}
\end{figure}


\begin{references}
\bibitem{qpnote} 
An excitation whose spectral line-width (inversely proportional to 
the lifetime) is longer than its energy (peak position) is, 
in the broadest sense,
called a quasiparticle, with no implication of a Fermi liquid picture.
\bibitem{EXPERIMENT}
The experiments were carried 
out using procedures and samples described previously 
\protect\cite{NATURE1}.
The doping level was controlled by varying oxygen stoichiometry, with
samples labeled by their onset $T_{c}$. Spectra were obtained
with a photon energy of 22 eV and a photon polarization
directed along the CuO bond direction. Spectra had energy resolutions 
(FWHM) of 15 - 25 meV with a momentum window of radius 0.045$\pi$/a. 
Energies are measured with respect to the chemical potential, determined 
using a polycrystalline Pt (or Au) reference in electrical contact with
the sample.
\end{references}
\end{document}